# Language-based Games

Adam Bjorndahl
Cornell University
Dept. Mathematics
Ithaca, NY 14853, USA
abjorndahl@math.cornell.edu

Joseph Y. Halpern
Cornell University
Dept. Computer Science
Ithaca, NY 14853, USA
halpern@cs.cornell.edu

Rafael Pass
Cornell University
Dept. Computer Science
Ithaca, NY 14853, USA
rafael@cs.cornell.edu

## ABSTRACT

We introduce *language-based games*, a generalization of *psychological games* [6] that can also capture *reference-dependent preferences* [7]. The idea is to extend the domain of the utility function to *situations*, maximal consistent sets in some language. The role of the underlying language in this framework is thus particularly critical. Of special interest are languages that can express only *coarse* beliefs [9]. Despite the expressive power of the approach, we show that it can describe games in a simple, natural way. Nash equilibrium and rationalizability are generalized to this setting; Nash equilibrium is shown not to exist in general, while the existence of rationalizable strategies is proved under mild conditions.

## Categories and Subject Descriptors

F.4.1 [**Mathematical Logic and Formal Languages**]: Mathematical Logic—*modal logic*; I.2.11 [**Artificial Intelligence**]: Distributed Artificial Intelligence—*multiagent systems*; J.4 [**Social and Behavioral Sciences**]: Economics

## 1. INTRODUCTION

In a classical, normal-form game, an *outcome* is a tuple of strategies, one for each player; intuitively, an outcome is just a record of which strategy each player chose to play. Players' preferences are formalized by utility functions defined on the set of all such outcomes. This framework thereby hard-codes the assumption that a player can prefer one state of the world to another only insofar as they differ in the outcome of the game.

Perhaps unsurprisingly, this model is too restrictive to account for a broad class of interactions that otherwise seem well-suited to a game-theoretic analysis. For example, one might wish to model players who feel guilt, wish to surprise their opponents, or are motivated by a desire to live up to what is expected of them. Work on *psychological game theory*, beginning with [6] and expanded in [3], is an enrichment of the classical setting meant to capture these kinds of preferences and motivations. In a similar vein, work on *reference-dependent preferences*, as developed in [7], formalizes phenomena such as loss-aversion by augmenting players' preferences with an additional sense of gain or loss derived by comparing the actual outcome to what was expected.

In both of these theories, the method of generalization takes the same basic form: the domain of the utility functions is enlarged to include not only the outcomes of the game, but also the beliefs of the players. The resulting structure may be fairly complex; for instance, in psychological game theory, since the goal is to model preferences that depend not only on beliefs about outcomes, but also beliefs about beliefs, beliefs about beliefs about beliefs, and so on, the domain of the utility functions is extended to include infinite hierarchies of beliefs.

The model we present in this paper, though motivated in part by a desire to capture belief-dependent preferences, is geared towards a much more general goal. Besides being expressive enough to subsume existing systems such as those described above, it establishes a general framework for modeling players with richer preferences. Moreover, it is equally capable of representing *impoverished* preferences, a canonical example of which are so-called "coarse beliefs" or "categorical thinking" [9]. More specifically, our formalism provides good practical and theoretical tools for handling beliefs as discrete rather than continuous objects, an advantage that is particularly relevant in the context of psychological effects in games.

Despite this expressive power, the system is easy to use: player preferences are represented in a simple and natural manner, narrowing the divide between intuition and formalism. As a preliminary illustration of some of these points, consider the following simple example.

*Example 1. A surprise proposal.* Alice and Bob have been dating for a while now, and Bob has decided that the time is right to pop the big question. Though he is not one for fancy proposals, he does want it to be a surprise. In fact, if Alice expects the proposal, Bob would prefer to postpone it entirely until such time as it might be a surprise. Otherwise, if Alice is not expecting it, Bob's preference is to take the opportunity.

We might summarize this scenario by the following table of payoffs for Bob:

|        | $p$ | $\neg p$ |
|--------|-----|----------|
| $B_A\, p$ | 0 | 1 |
| $\neg B_A\, p$ | 1 | 0 |

Table 1: The surprise proposal.

In this table, we denote Bob's two strategies, proposing and not proposing, as $p$ and $\neg p$, respectively, and use $B_A p$ (respectively, $\neg B_A p$) to denote that Alice is expecting (respectively, not expecting) the proposal.

Granted, whether or not Alice expects a proposal may be more than a binary affair: she may, for example, consider a proposal unlikely, somewhat likely, very likely, or certain. But there is good reason to think (see [9]) that an accurate model of her expectations stops here, with some small *finite* number $k$ of distinct "levels" of belief, rather than a continuum. Table 1, for simplicity, assumes that $k = 2$, though this is easily generalized to larger values.

Note that although Alice does not have a choice to make (formally, her strategy set is a singleton), she does have beliefs about which strategy Bob will choose. To represent Bob's preference for a surprise proposal, we must incorporate Alice's beliefs about Bob's choice of strategy into Bob's utility function. In psychological game theory, this is accomplished by letting $\alpha \in [0, 1]$ be the probability that Alice assigns to Bob proposing, and defining Bob's utility function $u_B$ in some simple way so that it is decreasing in $\alpha$ if Bob chooses to propose, and increasing in $\alpha$ otherwise:[1]

$$u_B(x, \alpha) = \begin{cases} 1 - \alpha & \text{if } x = p \\ \alpha & \text{if } x = \neg p. \end{cases}$$

The function $u_B$ agrees with the table at its extreme points if we identify $B_A p$ with $\alpha = 1$ and $\neg B_A p$ with $\alpha = 0$. Otherwise, for the infinity of other values that $\alpha$ may take between 0 and 1, $u_B$ yields a linear combination of the appropriate extreme points. Thus, in a sense, $u_B$ is a continuous approximation to a scenario that is essentially discrete.

We view Table 1 as *defining* Bob's utility. To coax an actual utility function from this table, let the variable $S$ denote a *situation*, which for the time being we can conceptualize as a collection of statements about the game; in this case, these include whether or not Bob is proposing, and whether or not Alice believes he is proposing. We then define

$$u_B(S) = \begin{cases} 0 & \text{if } p \in S \text{ and } B_A\, p \in S \\ 1 & \text{if } p \in S \text{ and } \neg B_A\, p \in S \\ 1 & \text{if } \neg p \in S \text{ and } B_A\, p \in S \\ 0 & \text{if } \neg p \in S \text{ and } \neg B_A\, p \in S. \end{cases}$$

In other words, Bob's utility is a function not merely of the outcome of the game ($p$ or $\neg p$), but of a more general object we are calling a "situation", and his utility in a given situation $S$ depends on his own actions combined with Alice's beliefs in exactly the manner prescribed by Table 1. As noted above, we may very well wish to refine our representation of Alice's state of surprise using more than two categories; indeed, we could allow a representation that permits continuous probabilities, as has been done in the literature. However, we will see that an "all-or-nothing" representation of belief is enough to capture some interesting and complex games. □

The central concept we develop in this paper is that of a *language-based game*, where utility is defined not on outcomes or the Cartesian product of outcomes with some other domain, but on *situations*. As noted, a situation can be conceptualized as a collection of statements about the game; intuitively, each statement is a description of something that might be relevant to player preferences, such as whether or not Alice believes that Bob will play a certain strategy. Of course, this notion crucially depends on just what counts as an admissible description. Indeed, the set of all admissible descriptions, which we refer to as the *underlying language* of the game, is a key component of our model. Since utility is defined on situations, and situations are sets of descriptions taken from the underlying language, a player's preferences can depend, in principle, on anything expressible in this language, and nothing more. Succinctly: players can prefer one state of the world to another if and only if they can *describe* the difference between the two, where "describe" here means "express in the underlying language".

Language-based games are thus parametrized by the underlying language: changing the language changes the game. The power and versatility of our approach derives in large part from this dependence. Consider, for example, an underlying language that contains only terms referring to players' strategies. Players' preferences, then, can depend only on the outcome of the game, as is the case classically. Thus classical game theory is recovered as a special case of the present work (see Sections 2.1 and 2.2 for details).

Enriching the underlying language allows for an expansion and refinement of player preferences; in this manner we are able to subsume, for example, work on psychological game theory and reference-dependent preferences, in addition to providing some uniformity to the project of defining new and further expansions of the classical base. By contrast, restricting the underlying language coarsens the domain of player preference; this provides a framework for modeling phenomena like coarse beliefs. A combination of these two approaches yields a theory of belief-dependent preferences incorporating coarse beliefs.

For the purposes of this paper, we focus primarily on belief-dependent preferences and coarseness, although in Example 6, we examine a simple scenario where a type of procrastination is represented by a minor extension of the underlying language. We make three major contributions. First, as noted, our system is easy to use in the sense that players' preferences are represented with a simple and uncluttered formalism; complex psychological phenomena can thus be captured in a direct and intuitive manner. Second, we provide a formal game-theoretic representation of coarse beliefs, and in so doing, expose an important insight: a discrete representation of belief, often conceptually and technically easier to work with than its continuous counterpart, is sufficient to capture psychological effects that have heretofore been modeled only in a continuous framework. Section 3 provides several examples that illustrate these points. Third, we provide novel equilibrium analyses that do not depend on the continuity of the expected utility function as in [6]. (Note that such continuity assumptions are at odds with our use of coarse beliefs.)

The rest of the paper is organized as follows. In the next section, we develop the basic apparatus needed to describe our approach. Section 3 presents a collection of examples intended to guide intuition and showcase the system. In Section 4, we show that there is a natural route by which solution concepts such as Nash equilibrium and rationalizability can be defined in our setting, and we address the question of existence. Proofs, more discussion, and further examples can be found in the full paper, which is available at http://www.cs.cornell.edu/home/halpern/papers/lbg.pdf.

---
[1]Technically, in [6], Bob's utility can only be a function of his own beliefs; this is generalized in [3] in the context of extensive-form games, but the approach is applicable to normal-form games as well.

## 2. FOUNDATIONS

### 2.1 Game forms and intuition

Much of the familiar apparatus of classical game theory is left untouched. A **game form** is a tuple $\Gamma = (N, (\Sigma_i)_{i \in N})$ where $N$ is a finite set of *players*, which for convenience we take to be the set $\{1, \ldots, n\}$, and $\Sigma_i$ is the set of *strategies available to player $i$*. Following standard notation, we set

$$\Sigma := \prod_{i \in N} \Sigma_i \quad \text{and} \quad \Sigma_{-i} := \prod_{j \neq i} \Sigma_j.$$

Elements of $\Sigma$ are called *outcomes* or *strategy profiles*; given $\sigma \in \Sigma$, we denote by $\sigma_i$ the $i$th component of the tuple $\sigma$, and by $\sigma_{-i}$ the element of $\Sigma_{-i}$ consisting of all but the $i$th component of $\sigma$.

Note that a game form does not come equipped with utility functions specifying the preferences of players over outcomes $\Sigma$. The utility functions we employ are defined on situations, which in turn are determined by the underlying language, so, before defining utility, we must first formalize these notions.

Informally, a *situation* is an exhaustive characterization of a given state of affairs using descriptions drawn from the underlying language. Assuming for the moment that we have access to a fixed "language", we might imagine a situation as being generated by simply listing all statements from that language that happen to be true of the world. Even at this intuitive level, it should be evident that the informational content of a situation is completely dependent on the expressiveness of the language. If, for example, the underlying language consists of exactly two descriptions, "It's raining" and "It's not raining", then there are only two situations:

{"It's raining"} and {"It's not raining"}.

Somewhat more formally, a situation $S$ is a set of formulas drawn from a larger pool of well-formed formulas, the underlying language. We require that $S$ include as many formulas as possible while still being consistent; we make this precise shortly.

The present formulation, informal though it is, is sufficient to allow us to capture a claim made in the introduction: any classical game can be recovered in our framework with the appropriate choice of underlying language. Specifically, let the underlying language be $\Sigma$, the set of all strategy profiles. Situations, in this case, are simply singleton subsets of $\Sigma$, as any larger set would contain distinct and thus intuitively contradictory descriptions of the outcome of the game. The set of situations can thus be identified with the set of outcomes, so a utility function defined on outcomes is readily identified with one defined on situations.

In this instance the underlying language, consisting solely of atomic, mutually incompatible formulas, is essentially structureless; one might wonder why call it a "language" at all, rather than merely a "set". Although, in principle, there are no restrictions on the kinds of objects we might consider as languages, it can be very useful to focus on those with some internal structure. This structure has two aspects: syntactic and semantic.

### 2.2 Syntax, semantics, and situations

The canonical form of syntactic structure in formal languages is *grammar*: a set of rules specifying how to compose well-formed formulas from atomic constituents. One of the best-known examples of a formal language generated by a grammar is the language of classical propositional logic.

Given a set $\Phi$ of *primitive propositions*, let $\mathcal{L}(\Phi)$ denote the propositional language based on $\Phi$, namely, the set of formulas that can be obtained by starting with $\Phi$ and closing off under $\neg$ and $\wedge$. (We can define $\vee$ and $\rightarrow$ from $\neg$ and $\wedge$ as usual.) Propositional logic is easily specialized to a game-theoretic setting. Given a game form $\Gamma = (N, (\Sigma_i)_{i \in N})$, let

$$\Phi_\Gamma = \{play_i(\sigma_i) \,:\, i \in N,\, \sigma_i \in \Sigma_i\},$$

where we read $play_i(\sigma_i)$ as "player $i$ is playing strategy $\sigma_i$". Then $\mathcal{L}(\Phi_\Gamma)$ is a language appropriate for reasoning about the strategies chosen by the players in $\Gamma$. We sometimes write $play(\sigma)$ as an abbreviation for $play_1(\sigma_1) \wedge \cdots \wedge play_n(\sigma_n)$.

*Semantics* provides a notion of truth. Recall that the semantics of classical propositional logic is given by *valuations* $v : \Phi \rightarrow \{\text{true}, \text{false}\}$. Valuations are extended to all formulas via the familiar truth tables for the logical connectives. Each valuation $v$ thereby generates a *model*, determining the truth values of every formula in $\mathcal{L}(\Phi)$. In the case of the language $\mathcal{L}(\Phi_\Gamma)$, we restrict this class of models to those corresponding to an outcome $\sigma \in \Sigma$; that is, we consider only valuation functions $v_\sigma$ defined by

$$v_\sigma(play_i(\sigma'_i)) = \text{true if and only if } \sigma_i = \sigma'_i.$$

More generally, we consider only a set $\mathcal{M}$ of *admissible models*: the ones that satisfy some restrictions of interest.

A set of formulas $F$ is said to be *satisfiable (with respect to a set $\mathcal{M}$ of admissible models)* if there is some model in $\mathcal{M}$ in which every formula of $F$ is true. An $\mathcal{L}(\Phi)$-*situation* is then defined to be a *maximal* satisfiable set of formulas (with respect to the admissible models of $\mathcal{L}(\Phi)$): that is, a satisfiable set with no proper superset that is also satisfiable. Situations correspond to admissible models: a situation just consists of all the formulas true in some admissible model. Let $\mathcal{S}(\mathcal{L}(\Phi))$ denote the set of $\mathcal{L}(\Phi)$-situations. It is not difficult to see that $\mathcal{S}(\mathcal{L}(\Phi_\Gamma))$ can be identified with the set $\Sigma$ of outcomes.

Having illustrated some of the principle concepts of our approach in the context of propositional logic, we now present the definitions in complete generality. Let $\mathcal{L}$ be a language with an associated semantics, that is, a set of admissible models providing a notion of truth. We often use the term "language" to refer to a set of well-formed formulas together with a set of admissible models (this is sometimes called a "logic"). An $\mathcal{L}$-**situation** is a maximal satisfiable set of formulas from $\mathcal{L}$. Denote by $\mathcal{S}(\mathcal{L})$ the set of $\mathcal{L}$-situations. A game form $\Gamma$ is extended to an $\mathcal{L}$-**game** by adding utility functions $u_i : \mathcal{S}(\mathcal{L}) \rightarrow \mathbb{R}$, one for each player $i \in N$. $\mathcal{L}$ is called the **underlying language**; we omit it as a prefix when it is safe to do so.

If we view $\Gamma$ as an $\mathcal{L}(\Phi_\Gamma)$-game, the players' utility functions are essentially defined on $\Sigma$, so an $\mathcal{L}(\Phi_\Gamma)$-game is really just a classical game based on $\Gamma$. As we saw in Section 2.1, this class of games can also be represented with the completely structureless language $\Sigma$. This may well be sufficient, especially in cases where all we care about are two or three formulas. However, having a structured underlying language makes it easier to analyze the much broader class of psychological games.

A psychological game is just like a classical game except that players' preferences can depend not only on what strategies are played, but also on what beliefs are held. While $\mathcal{L}(\Phi_\Gamma)$ is appropriate for reasoning about strategies, it cannot express anything about beliefs, so our first step is to define a richer language. Fortunately, we have at our disposal a host of candidates well-equipped for this task, namely those languages associated with epistemic logics.

Fix a game form $\Gamma = (N, (\Sigma_i)_{i \in N})$, and let $\mathcal{L}_B(\Phi_\Gamma)$ be the language obtained by starting with the primitive propositions in $\Phi_\Gamma$ and closing off under conjunction, negation, and the modal operators $B_i$, for $i \in N$. We read $B_i \varphi$ as "player $i$ believes $\varphi$". Intuitively, this is a language for reasoning about the beliefs of the players and the strategy profiles being used.

We give semantics to $\mathcal{L}_B(\Phi_\Gamma)$ using Kripke structures, as usual. But for many examples of interest, understanding the (completely standard, although somewhat technical) details is not necessary. Example 1 was ultimately analyzed as an $\mathcal{L}_B(\Phi_\Gamma)$-game, despite the fact that we had not even defined the syntax of this language at the time, let alone its semantics. Section 3 provides more illustrations of this point.

A **$\Gamma$-structure** is a tuple $M = (\Omega, \vec{s}, Pr_1, \ldots, Pr_n)$ satisfying the following conditions:

(P1) $\Omega$ is a nonempty topological space;

(P2) each $Pr_i$ assigns to each $\omega \in \Omega$ a probability measure $Pr_i(\omega)$ on $\Omega$;

(P3) $\omega' \in Pr_i[\omega] \Rightarrow Pr_i(\omega') = Pr_i(\omega)$, where $Pr_i[\omega]$ abbreviates $supp(Pr_i(\omega))$, the support of the probability measure;

(P4) $\vec{s}: \Omega \to \Sigma$ satisfies $Pr_i[\omega] \subseteq \{\omega' : s_i(\omega') = s_i(\omega)\}$, where $s_i(\omega)$ denotes player $i$'s strategy in the strategy profile $\vec{s}(\omega)$.

These conditions are standard for KD45 belief logics in a game-theoretic setting [1]. The set $\Omega$ is called the **state space**. Conditions (P1) and (P2) set the stage to represent player $i$'s beliefs in state $\omega \in \Omega$ as the probability measure $Pr_i(\omega)$ over the state space itself. Condition (P3) says essentially that players are sure of their own beliefs. The function $\vec{s}$ is called the **strategy function**, assigning to each state a strategy profile that we think of as the strategies that the players are playing at that state. Condition (P4) thus asserts that each player is sure of his own strategy. The language $\mathcal{L}_B(\Phi_\Gamma)$ can be interpreted in any $\Gamma$-structure $M$ via the strategy function, which induces a valuation $\llbracket \cdot \rrbracket_M : \mathcal{L}_B(\Phi_\Gamma) \to 2^\Omega$ defined recursively by:

$$\begin{array}{rcl}
\llbracket play_i(\sigma_i) \rrbracket_M & := & \{\omega \in \Omega : s_i(\omega) = \sigma_i\} \\
\llbracket \varphi \wedge \psi \rrbracket_M & := & \llbracket \varphi \rrbracket_M \cap \llbracket \psi \rrbracket_M \\
\llbracket \neg \varphi \rrbracket_M & := & \Omega - \llbracket \varphi \rrbracket_M \\
\llbracket B_i \varphi \rrbracket_M & := & \{\omega \in \Omega : Pr_i[\omega] \subseteq \llbracket \varphi \rrbracket_M\}.
\end{array}$$

Thus, the Boolean connectives are interpreted classically, and $B_i \varphi$ holds at state $\omega$ just in case all the states in the support of $Pr_i(\omega)$ are states where $\varphi$ holds.

Pairs of the form $(M, \omega)$, where $M = (\Omega, \vec{s}, \vec{Pr})$ is a $\Gamma$-structure and $\omega \in \Omega$, play the role of admissible models for the language $\mathcal{L}_B(\Phi_\Gamma)$. Given $\varphi \in \mathcal{L}_B(\Phi_\Gamma)$, we sometimes write $(M, \omega) \models \varphi$ or just $\omega \models \varphi$ instead of $\omega \in \llbracket \varphi \rrbracket_M$, and say that $\omega$ **satisfies** $\varphi$ or $\varphi$ is **true at** $\omega$; we write $M \models \varphi$ and say that $\varphi$ is **valid in** $M$ if $\llbracket \varphi \rrbracket_M = \Omega$. We say that $\varphi$ is **satisfiable** if for some state $\omega$ in some $\Gamma$-structure $M$ (i.e., for some admissible model), $\omega \models \varphi$. Given $F \subseteq \mathcal{L}_B(\Phi_\Gamma)$, we write $\omega \models F$ if for all $\varphi \in F$, $\omega \models \varphi$; we say that $F$ is satisfiable if for some state $\omega$ in some $M$, $\omega \models F$.

With this notion of satisfiability, we gain access to the class of $\mathcal{L}_B(\Phi_\Gamma)$-games, where utility is defined on $\mathcal{L}_B(\Phi_\Gamma)$-situations, namely, maximal satisfiable subsets of $\mathcal{L}_B(\Phi_\Gamma)$. In particular, we can extend any game form $\Gamma$ to an $\mathcal{L}_B(\Phi_\Gamma)$-game, a setting in which players' preferences can depend, in principle, on anything describable in the language $\mathcal{L}_B(\Phi_\Gamma)$.

It is not hard to show that when there is more than one player, $\mathcal{S}(\mathcal{L}_B(\Phi_\Gamma))$ is uncountable. A utility function $u_i : \mathcal{S}(\mathcal{L}_B(\Phi_\Gamma)) \to \mathbb{R}$ can therefore be quite complicated indeed. We will frequently be interested in representing preferences that are much simpler. For instance, though the surprise proposal scenario presented in Example 1 can be viewed as an $\mathcal{L}_B(\Phi_\Gamma)$-game, Bob's utility $u_B$ does not depend on any situation as a whole, but rather is determined by a few select formulas. This motivates the following general definition, identifying a particularly easy to understand and well-behaved subclass of games.

Fix a language $\mathcal{L}$. A function $u : \mathcal{S}(\mathcal{L}) \to \mathbb{R}$ is called **finitely specified** if there is a finite set of formulas $F \subset \mathcal{L}$ and a function $f : F \to \mathbb{R}$ such that every situation $S \in \mathcal{S}(\mathcal{L})$ contains exactly one formula from $F$, and whenever $\varphi \in S \cap F$, $u(S) = f(\varphi)$. In other words, the value of $u$ depends only on the formulas in $F$. Thus $u$ is finitely specified if and only if it can be written in the form

$$u(S) = \begin{cases} a_1 & \text{if } \varphi_1 \in S \\ \vdots & \vdots \\ a_k & \text{if } \varphi_k \in S, \end{cases}$$

for some $a_1, \ldots, a_k \in \mathbb{R}$ and $\varphi_1, \ldots, \varphi_k \in \mathcal{L}$.

A language-based game is called finitely specified if each player's utility function is. Many games of interest are finitely specified. In a finitely specified game, we can think of a player's utility as being a function of the finite set $F$; indeed, we can think of the underlying language as being the structureless "language" $F$ rather than $\mathcal{L}$.

## 3. EXAMPLES

We now give a few examples to exhibit both the simplicity and the expressive power of language-based games; more examples are given in the full paper. Since we focus on the language $\mathcal{L}_B(\Phi_\Gamma)$, we write $\mathcal{S}$ to abbreviate $\mathcal{S}(\mathcal{L}_B(\Phi_\Gamma))$.

Note that there is a unique strategy that player $i$ uses in a situation $S \in \mathcal{S}$; it is the strategy $\sigma_i$ such that $play_i(\sigma_i) \in S$. When describing the utility of a situation, it is often useful to extract this strategy; therefore, we define $\rho_i : \mathcal{S} \to \Sigma_i$ implicitly by the requirement $play_i(\rho_i(S)) \in S$. It is easy to check that $\rho_i$ is well-defined.

*Example 2. Indignant altruism.* Alice and Bob sit down to play a classic game of prisoner's dilemma, with one twist:

neither wishes to live up to low expectations. Specifically, if Bob expects the worst of Alice (i.e. expects her to defect), then Alice, indignant at Bob's opinion of her, prefers to cooperate. Likewise for Bob. On the other hand, in the absence of such low expectations from their opponent, each will revert to their classical, self-serving behaviour.

The standard prisoner's dilemma is summarized in Table 2:

|   | c | d |
|---|---|---|
| c | (3,3) | (0,5) |
| d | (5,0) | (1,1) |

**Table 2: The classical prisoner's dilemma.**

Let $u_A$, $u_B$ denote the two players' utility functions according to this table, and let $\Gamma$ denote the game form obtained by throwing away these functions: $\Gamma = (\{A, B\}, \Sigma_A, \Sigma_B)$ where $\Sigma_A = \Sigma_B = \{\mathsf{c}, \mathsf{d}\}$. We wish to define an $\mathcal{L}_B(\Phi_\Gamma)$-game that captures the given scenario; to do so we must define new utility functions on $\mathcal{S}$. Informally, if Bob is sure that Alice will defect, then Alice's utility for defecting is $-1$, regardless of what Bob does, and likewise reversing the roles of Alice and Bob; otherwise, utility is determined exactly as it is classically.

Formally, we simply define $u'_A : \mathcal{S} \to \mathbb{R}$ by

$$u'_A(S) = \begin{cases} -1 & \text{if } play_A(\mathsf{d}) \in S \text{ and} \\ & B_B\, play_A(\mathsf{d}) \in S \\ u_A(\rho_A(S), \rho_B(S)) & \text{otherwise,} \end{cases}$$

and similarly for $u'_B$.

Intuitively, cooperating is rational for Alice if she thinks that Bob is sure she will defect, since cooperating in this case would yield a minimum utility of 0, whereas defecting would result in a utility of $-1$. On the other hand, if Alice thinks that Bob is *not* sure she'll defect, then since her utility in this case would be determined classically, it is rational for her to defect, as usual.

This game has much in common with the surprise proposal of Example 1: in both games, the essential psychological element is the desire to surprise another player. Perhaps unsurprisingly, when players wish to surprise their opponents, *Nash equilibria* fail to exist—even mixed strategy equilibria. Although we have not yet defined Nash equilibrium in our setting, the classical intuition is wholly applicable: a Nash equilibrium is a state of play where players are happy with their choice of strategies *given accurate beliefs about what their opponents will choose*. But there is a fundamental tension between a state of play where everyone has accurate beliefs, and one where some player successfully surprises another.

We show formally in Section 4.2 that this game has no Nash equilibrium. On the other hand, players can certainly best-respond to their beliefs, and the corresponding iterative notion of *rationalizability* finds purchase here. In Section 4.3 we will import this solution concept into our framework and show that every strategy for the indignant altruist is rationalizable.  □

*Example 3. A deeply surprising proposal.* Bob hopes to propose to Alice, but she wants it to be a surprise. He knows that she would be upset if it were not a surprise, so he would prefer not to propose if Alice so much as suspects it. Worse (for Bob), even if Alice does not suspect a proposal, if she suspects that Bob thinks she does, then she will also be upset, since in this case a proposal would indicate Bob's willingness to disappoint her. Of course, like the giant tortoise on whose back the world rests, this reasoning continues "all the way down"...

This example is adapted from a similar example given in [6]; in that example, the man is considering giving a gift of flowers, but rather than hoping to surprise the recipient, his goal is the exact opposite: to get her flowers just in case she *is* expecting them. Of course, the notion of "expectation" employed, both in their example and ours, is quite a bit more complicated than the usual sense of the word, involving arbitrarily deeply nested beliefs.

Nonetheless, it is relatively painless to represent Bob's preferences in the language $\mathcal{L}_B(\Phi_\Gamma)$, where $\Gamma = (\{A, B\}, \{\cdot\}, \{p, q\})$ and $p$ and $q$ stand for Bob's strategies of proposing and not proposing, respectively (Alice has no decision to make, so her strategy set is a singleton). For convenience, we use the symbol $P_i$ to abbreviate $\neg B_i \neg$. Thus $P_i \varphi$ holds just in case player $i$ is not sure that $\varphi$ is false; this will be our gloss for Alice "so much as suspecting" a proposal. Define $u_B : \mathcal{S} \to \mathbb{R}$ by

$$u_B(S) = \begin{cases} 1 & \text{if } play_B(p) \in S \text{ and} \\ & (\forall k \in \mathbb{N})[P_A(P_B P_A)^k play_B(p) \notin S] \\ 1 & \text{if } play_B(q) \in S \text{ and} \\ & (\exists k \in \mathbb{N})[P_A(P_B P_A)^k play_B(p) \in S] \\ 0 & \text{otherwise,} \end{cases}$$

where $(P_B P_A)^k$ is an abbreviation for $P_B P_A \cdots P_B P_A$ ($k$ times). In other words, proposing yields a higher utility for Bob in the situation $S$ if and only if *none* of the formulas in the infinite family $\{P_A(P_B P_A)^k play_B(p) : k \in \mathbb{N}\}$ occur in $S$.

As in Examples 1 and 2, and in general when a player desires to surprise an opponent, it is not difficult to convince oneself informally that this game admits no Nash equilibrium. Moreover, in this case the infinitary nature of Bob's desire to "surprise" Alice has an even stronger effect: no strategy for Bob is even *rationalizable* (see Section 4.3).  □

*Example 4. Pay raise.* Bob has been voted employee of the month at his summer job, an honour that comes with a slight increase (up to $1) in his per-hour salary, at the discretion of his boss, Alice. Bob's happiness is determined in part by the raw value of the bump he receieves in his wages, and in part by the sense of gain or loss he feels by comparing the increase Alice grants him with the minimum increase he expected to get. Alice, for her part, wants Bob to be happy, but this desire is balanced by a desire to save company money.

As usual, we first fix a game form that captures the players and their available strategies. Let $\Gamma = (\{A, B\}, \Sigma_A, \{\cdot\})$, where $\Sigma_A = \{s_0, s_1, \ldots, s_{100}\}$ and $s_k$ represents an increase of $k$ cents to Bob's per-hour salary (Bob has no choice to make, so his strategy set is a singleton). Notice that, in contrast to the other examples we have seen thus far, in this game Bob's preferences depend on his *own* beliefs rather than the beliefs of his opponent. Broadly speaking, this is an example of *reference-dependent preferences*: Bob's utility is determined in part by comparing the actual outcome of the game to some "reference level"—in this case, the minimum expected raise. This game also has much in common with a scenario described in [3], in which a player Abi wishes to tip her taxi driver exactly as much as he expects to be tipped, but no more.

Define $u_B : \mathcal{S} \to \mathbb{R}$ by

$$u_B(S) = k + (k - r),$$

where $k$ is the unique integer such that $play_A(s_k) \in S$, and

$$r := \min\{r' \ : \ P_B \, play_A(s_{r'}) \in S\}.$$

Observe that $r$ is completely determined by Bob's beliefs: it is the lowest raise he considers it possible that Alice will grant him. We think of the first summand $k$ as representing Bob's happiness on account of receiving a raise of $k$ cents per hour, while the second summand $k - r$ represents his sense of gain or loss depending on how reality compares to his lowest expectations.

Note that the value of $r$ (and $k$) is encoded in $S$ via a finite formula, so we could have written the definition of $u_B$ in a fully expanded form where each utility value is specified by the presense of a formula in $S$. For instance, the combination $k = 5$, $r = 2$ corresponds to the formula

$$play_A(s_5) \wedge P_B \, play_A(s_2) \wedge \neg(P_B \, play_A(s_0) \vee P_B \, play_A(s_1)),$$

which therefore determines a utility of 8.

Of course, it is just as easy to replace the minimum with the maximum in the above definition (perhaps Bob feels entitled to the most he considers it possible he might get), or even to define the reference level $r$ as some more complicated function of Bob's beliefs. The quantity $k - r$ representing Bob's sense of gain or loss is also easy to manipulate. For instance, given $\alpha, \beta \in \mathbb{R}$ we might define a function $f : \mathbb{R} \to \mathbb{R}$ by

$$f(x) = \begin{cases} \alpha x & \text{if } x \geq 0 \\ \beta x & \text{if } x < 0, \end{cases}$$

and set

$$u'_B(S) = k + f(k - r),$$

where $k$ and $r$ are determined as above. Choosing, say, $\alpha = 1$ and $\beta > 1$ results in Bob's utility $u'_B$ incorporating *loss aversion*: Bob is more upset by a relative loss than he is elated by a same-sized relative gain. These kinds of issues are discussed in [7]; in the full paper we analyze a central example from this paper in detail.

Turning now to Alice's preferences, we are faced with a host of modeling choices. Perhaps Alice wishes to grant Bob the smallest salary increase he expects but nothing more. We can capture this by defining $u_A : \mathcal{S} \to \mathbb{R}$ by

$$u_A(S) = -|k - r|,$$

where $k$ and $r$ are as above. Or perhaps we wish to represent Alice as feeling some fixed sense of guilt if she undershoots, while her disutility for overshooting depends on whether she merely exceeded Bob's lowest expectations, or in fact exceeded even his highest expectations:

$$u'_A(S) = \begin{cases} -25 & \text{if } k < r \\ r - k & \text{if } r \leq k < R \\ r - R + 2(R - k) & \text{if } k \geq R, \end{cases}$$

where

$$R := \max\{R' \ : \ P_B \, play_A(s_{R'}) \in S\}.$$

Or perhaps Alice's model of Bob's happiness is sophisticated enough to *include* his sensations of gain and loss, so that, for example,

$$u''_A(S) = u_B(S) - \delta k,$$

where $\delta$ is some scaling factor. Clearly the framework is rich enough to represent many possibilities. □

*Example 5. Preparing for a roadtrip.* Alice has two tasks to accomplish before embarking on a cross-country roadtrip: she needs to buy a suitcase, and she needs to buy a car.

Here we sketch a simple decision-theoretic scenario in a language-based framework. We choose the underlying language in such a way as to capture two well-known "irrationalities" of consumers. First, consumers often evaluate prices in a discontinuous way, behaving, for instance, as if the difference between \$299 and \$300 is more substantive than the difference between \$300 and \$301. Second, consumers who are willing to put themselves out (for example, drive an extra 5 kilometers) to save \$50 on a \$300 purchase are often not willing to drive that same extra distance to save the same amount of money on a \$20,000 purchase.

We do not claim a completely novel analysis; rather, we aim to show how naturally a language-based approach can account for these kinds of issues.

Both of the irrationalities described above can be captured by assuming a certain kind of coarseness, specifically, that the language over which Alice forms preferences does not describe prices with infinite precision. For example, we might assume that the language includes as primitive propositions terms of the form $p_Q$, where $Q$ ranges over a given partition of the real line. We might further suppose that this partition has the form

$$\cdots \cup [280, 290) \cup [290, 300) \cup [300, 310) \cup \cdots,$$

at least around the \$300 mark. Any utility function defined over such a language cannot distinguish prices that fall into the same partition. Thus, in the example above, Alice would consider the prices \$300 and \$301 to be effectively the same as far as her preferences are concerned. At the borderline between cells of the partition, however, there is the potential for a "jump": we might reasonably model Alice as prefering a situation where $p_{[290,300)}$ holds to one where $p_{[300,310)}$ holds. A smart retailer, therefore, should set their price to be at the higher end of a cell of the consumers' partition.

To capture the second irrationality discussed above, it suffices to assume that the partition that determines the underlying language is not only coarse, but is coarser for higher prices. For example, around the \$20,000 mark, we might suppose that the partition has the form

$$\cdots \cup [19000, 19500) \cup [19500, 20000) \cup [20000, 20500) \cup \cdots.$$

In this case, while Alice may prefer a price of \$300 to a price of \$350, she cannot prefer a price of \$20,000 to a price of \$20,050, because that difference cannot be described in the underlying language. This has a certain intuitive appeal: the higher numbers get (or, more generally, the further removed something is, in space or time or abstraction), the more you "ballpark" it—the less precise your language is in describing it. Indeed, psychological experiments have demonstrated that Weber's law[2], traditionally applied to physical stimuli, finds purchase in the realm of numerical perception: larger numbers are subjectively harder to discriminate from one another [8; 11]. Our choice of underlying language represents this phenomenon simply, while exhibiting its explanatory power. □

---

[2] Weber's law asserts that the minimum difference between two stimuli necessary for a subject to discriminate between them increases as the magnitude of the stimuli increases.

*Example 6. Returning a library book.* Alice has learned that a book she borrowed from the library is due back tomorrow. As long as she returns it by tomorrow, she'll avoid a late fee; returning it today, however, is mildly inconvenient.

Here we make use of an extremely simple example to illustrate how to model an ostensibly dynamic scenario in a normal-form framework by employing a suitable underlying language. The idea is straightforward: Alice has a choice to make *today*, but how she feels about it depends on what she might do tomorrow. Specifically, if she returns the library book tomorrow, then she has no reason to feel bad about not returning it today. Since the future has yet to be determined, we model Alice's preferences as depending on what action she takes in the present together with what she *expects* to do in the future.

Let $\Gamma = (A, \{\text{return}, \text{wait}\})$ be a game form representing Alice's two current options, and set $\Phi'_\Gamma := \Phi_\Gamma \cup \{\text{tomorrow}\}$; thus $\Phi'_\Gamma$ is the usual set of primitive propositions (representing strategies) together with a single new addition, tomorrow, read "Alice will return the book tomorrow".

An $\mathcal{L}_B(\Phi'_\Gamma)$-game allows us to specify Alice's utility in a manner consistent with the intuition given above. In particular, we can define $u_A : \mathcal{S}(\mathcal{L}_B(\Phi'_\Gamma)) \to \mathbb{R}$ by

$$u_A(S) = \begin{cases} -1 & \text{if } play_A(\text{return}) \in S \\ 1 & \text{if } play_A(\text{wait}) \wedge B_A\text{tomorrow} \in S \\ -5 & \text{otherwise,} \end{cases}$$

so Alice prefers to wait if she expects to return the book tomorrow, and to return the book today otherwise.

In this example, Alice's utility depends on her beliefs, as it does in psychological game theory. Unlike psychological game theory, however, her utility depends on her beliefs about features of the world aside from which strategies are being played. This is a natural extension of the psychological framework in a language-based setting.

This example also hints at another interesting application of language-based games. A careful look at the language $\mathcal{L}_B(\Phi'_\Gamma)$ reveals an oddity: as far as the semantics are concerned, $play_A(\text{return})$ and tomorrow are independent primitive propositions, despite being intuitively contradictory. Of course, this can be rectified easily enough: we can simply insist in the semantics that whenever $play_A(\text{return})$ holds at a state, tomorrow does not. But in so doing, we have introduced a further complexity: the strategy that Alice chooses now determines more about the situation than merely the fact of which strategy she has chosen.

This observation reveals the need for a good theory of counterfactuals. After all, it is not just the true state of the world that must satisfy the semantic contraints we impose, but also the counterfactual situations we consider when determining whether or not a player is behaving rationally. In Section 4.1, we give a formal treatment of rationality in $\mathcal{L}_B(\Phi_\Gamma)$-games that skirts this issue; however, we believe that a more substantive treatment of counterfactual reasoning in games is both important and interesting, and that the present framework is a promising setting in which to develop such a theory.

Returning to the example at hand, we might emphasize the new element of "control" Alice has by providing her with explicit mechanisms of influencing her own beliefs about tomorrow. For example, perhaps a third strategy is available to her, remind, describing a state of affairs where she keeps the book but places it on top of her keys, thus decreasing the likelihood that she will forget to take it when she leaves the next day.

More generally, this simple framework allows us to model *commitment devices* [5]: we can represent players who rationally choose to perform certain actions (like buying a year-long gym membership, or throwing away their "fat jeans") not because these actions benefit them immediately, but because they make it subjectively more likely that the player will perform certain other desirable actions in the future (like going to the gym regularly, or sticking with a diet) that might otherwise be neglected. In a similar manner, we can succinctly capture *procrastination*: if, for example, you believe that you will quit smoking tomorrow, then the health benefits of quitting today instead might seem negligible—so negligible, in fact, that quitting immediately may seem pointless, even foolish. Of course, believing you will do something tomorrow is not the same thing as actually doing it when tomorrow comes, thus certain tasks may be delayed repeatedly. □

## 4. SOLUTION CONCEPTS

A number of important concepts from classical game theory, such as *Nash equilibrium* and *rationalizability*, have been completely characterized epistemically, using $\Gamma$-structures. In $\mathcal{L}_B(\Phi_\Gamma)$-games (or, more generally, in language-based games where the language includes belief), we can use the epistemic characterizations as the *definitions* of these solution concepts. This yields natural definitions that generalize those of classical game theory. We begin by defining rationality in our setting.

### 4.1 Rationality

We call a player $i$ *rational* if he is best-responding to his beliefs: the strategy $\sigma_i$ he is using must yield an expected utility that is at least as good as any other strategy $\sigma'_i$ he could play, given his beliefs. In classical game theory, the meaning of this statement is quite clear. Player $i$ has beliefs about the strategy profiles $\sigma_{-i}$ used by the other players. This makes it easy to compute what $i$'s payoffs would be if he were to use some other strategy $\sigma'_i$: since $i$'s utility just depends on the strategy profile being used, we simply replace $\sigma_i$ by $\sigma'_i$ in these strategy profiles, and compute the new expected utility. Thus, for example, in a two-player game, if player 1 places probability $1/2$ on the two strategies $\sigma_2$ and $\sigma'_2$ for player 2, then his expected utility playing $\sigma_1$ is $(u_1(\sigma_1, \sigma_2) + u_1(\sigma_1, \sigma'_2))/2$, while his expected utility if he were to play $\sigma'_1$ is $(u_1(\sigma'_1, \sigma_2) + u_1(\sigma'_1, \sigma'_2))/2$.

We make use of essentially the same approach in language-based games. Let $(\Gamma, (u_i)_{i \in N})$ be an $\mathcal{L}_B(\Phi_\Gamma)$-game and fix a $\Gamma$-structure $M = (\Omega, \vec{s}, \vec{Pr})$. Observe that for each $\omega \in \Omega$ and each $i \in N$, there is a unique $\mathcal{L}_B(\Phi_\Gamma)$-situation $S$ such that $\omega \models S$; we denote this situation by $S(M, \omega)$ or just $S(\omega)$ when the $\Gamma$-structure is clear from context.

If $play_i(\sigma_i) \in S(\omega)$, then given $\sigma'_i \in \Sigma_i$ we might naïvely let $S(\omega/\sigma'_i)$ denote the set $S(\omega)$ with the formula $play_i(\sigma_i)$ replaced by $play_i(\sigma'_i)$, and define $\hat{u}_i(\sigma'_i, \omega)$, the utility that $i$ would get if he played $\sigma'_i$ in state $\omega$, as $u_i(S(\omega/\sigma'_i))$. Un-fortunately, $u_i$ is not necessarily defined on $S(\omega/\sigma_i')$, since it is not the case in general that this set is satisfiable; indeed, $S(\omega/\sigma_i')$ is satisfiable if and only if $\sigma_i' = \sigma_i$. This is because other formulas in $S(\omega)$, for example the formula $B_i\, play_i(\sigma_i)$, logically imply the formula $play_i(\sigma_i)$ that was removed from $S(\omega)$ (recall that our semantics insist that every player is sure of their own strategy). With a more careful construction of the "counterfactual" set $S(\omega/\sigma_i')$, however, we can obtain a definition of $\hat{u}_i$ that makes sense.

A formula $\varphi \in \mathcal{L}_B(\Phi_\Gamma)$ is called *i*-**independent** if for each $\sigma_i \in \Sigma_i$, every occurrence of $play_i(\sigma_i)$ in $\varphi$ falls within the scope of some $B_j$, $j \neq i$. Intuitively, an *i*-independent formula describes a proposition that is independent of player *i*'s choice of strategy, such as another player's strategy, another player's beliefs, or even player *i*'s beliefs about the other players; on the other hand, player *i*'s beliefs about his own choices are excluded from this list, as they are assumed to always be accurate, and thus dependent on those choices. Given $S \in \mathcal{S}$, set

$$\rho_{-i}(S) = \{\varphi \in S \,:\, \varphi \text{ is } i\text{-independent}\}.[3]$$

Let $\mathcal{S}_{-i}$ denote the image of $\mathcal{S}$ under $\rho_{-i}$. Elements of $\mathcal{S}_{-i}$ are called *i*-**situations**; intuitively, they are complete descriptions of states of affairs that are out of player *i*'s control. Informally, an *i*-situation $S_{-i} \in \mathcal{S}_{-i}$ determines everything about the world (expressible in the language) *except* what strategy player *i* is employing. This is made precise in Proposition 1. Recall that $\rho_i(S)$ denotes the (unique) strategy that *i* plays in $S$, so $play_i(\rho_i(S)) \in S$.

PROPOSITION 1. *For each $i \in N$, the map $\vec{\rho}_i : \mathcal{S} \to \Sigma_i \times \mathcal{S}_{-i}$ defined by $\vec{\rho}_i(S) = (\rho_i(S), \rho_{-i}(S))$ is a bijection.*

This identification of $\mathcal{S}$ with the set of pairs $\Sigma_i \times \mathcal{S}_{-i}$ provides a well-defined notion of what it means to alter player *i*'s strategy in a situation $S$ "without changing anything else". By an abuse of notation, we write $u_i(\sigma_i, S_{-i})$ to denote $u_i(S)$ where $S$ is the unique situation corresponding to the pair $(\sigma_i, S_{-i})$, that is, $\vec{\rho}_i(S) = (\sigma_i, S_{-i})$. Observe that for each state $\omega \in \Omega$ and each $i \in N$ there is a unique set $S_{-i} \in \mathcal{S}_{-i}$ such that $\omega \models S_{-i}$. We denote this set by $S_{-i}(M, \omega)$, or just $S_{-i}(\omega)$ when the $\Gamma$-structure is clear from context. Then the utility functions $u_i$ induce functions $\hat{u}_i : \Sigma_i \times \Omega \to \mathbb{R}$ defined by

$$\hat{u}_i(\sigma_i, \omega) = u_i(\sigma_i, S_{-i}(\omega)).$$

As in the classical case, we can view the quantity $\hat{u}_i(\sigma_i, \omega)$ as the utility that player *i* would have if he were to play $\sigma_i$ at state $\omega$. It is easy to see that this generalizes the classical approach in the sense that it agrees with the classical definition when the utility functions $u_i$ depend only on the outcome.

For each $i \in N$, let $EU_i : \Sigma_i \times \Omega \to \mathbb{R}$ be the expected utility of playing $\sigma_i$ according to player *i*'s beliefs at $\omega$. Formally:

$$EU_i(\sigma_i, \omega) = \int_\Omega \hat{u}_i(\sigma_i, \omega')\, dPr_i(\omega);$$

---

[3] As (quite correctly) pointed out by an anonymous reviewer, this notation is not standard, since $\rho_{-i}$ is not a profile of functions of the type $\rho_i$. Nonetheless, we feel it is appropriate in the sense that, while $\rho_i$ extracts from a given situation player *i*'s strategy, $\rho_{-i}$ extracts "all the rest" (cf. Proposition 1), the crucial difference here being that this includes far more than just the strategies of the other players.

when $\Omega$ is finite, this reduces to

$$EU_i(\sigma_i, \omega) = \sum_{\omega' \in \Omega} \hat{u}_i(\sigma_i, \omega') \cdot Pr_i(\omega)(\omega').$$

Define $BR_i : \Omega \to 2^{\Sigma_i}$ by

$$BR_i(\omega) = \{\sigma_i \in \Sigma_i \,:\, (\forall \sigma_i' \in \Sigma_i)[EU_i(\sigma_i, \omega) \geq EU_i(\sigma_i', \omega)]\};$$

thus $BR_i(\omega)$ is the set of *best-reponses* of player *i* to his beliefs at $\omega$, that is, the set of strategies that maximize his expected utility.

With this apparatus in place, we can expand the underlying language to incorporate *rationality* as a formal primitive. Let

$$\Phi_\Gamma^{rat} \coloneqq \Phi_\Gamma \cup \{RAT_i \,:\, i \in N\},$$

where we read $RAT_i$ as "player *i* is rational". We also employ the syntactic abbreviation $RAT \equiv RAT_1 \wedge \cdots \wedge RAT_n$. Intuitively, $\mathcal{L}_B(\Phi_\Gamma^{rat})$ allows us to reason about whether or not players are being rational with respect to their beliefs and preferences.

We wish to interpret rationality as expected utility maximization. To this end, we extend the valuation function $[\![\cdot]\!]_M$ to $\mathcal{L}_B(\Phi_\Gamma^{rat})$ by

$$[\![RAT_i]\!]_M \;\coloneqq\; \{\omega \in \Omega \,:\, s_i(\omega) \in BR_i(\omega)\}.$$

Thus $RAT_i$ holds at state $\omega$ just in case the strategy that player *i* is playing at that state, $s_i(\omega)$, is a best-response to his beliefs.

## 4.2 Nash equilibrium

Having formalized rationality, we are in a position to draw on work that characterizes solutions concepts in terms of $RAT$.

Let $\Gamma = (N, (\Sigma_i)_{i \in N})$ be a game form in which each set $\Sigma_i$ is finite, and let $\Delta(\Sigma_i)$ denote the set of all probability measures on $\Sigma_i$. Elements of $\Delta(\Sigma_i)$ are the **mixed strategies** of player *i*. Given a *mixed strategy profile*

$$\mu = (\mu_1, \ldots, \mu_n) \in \Delta(\Sigma_1) \times \cdots \times \Delta(\Sigma_n),$$

we define a $\Gamma$-structure $M_\mu$ that, in a sense made precise below, captures "equilibrium play" of $\mu$ and can be used to determine whether or not $\mu$ constitutes a Nash equilibrium. Set

$$\Omega_\mu = supp(\mu_1) \times \cdots \times supp(\mu_n) \subseteq \Sigma_1 \times \cdots \times \Sigma_n.$$

Define a probability measure $\pi$ on $\Omega_\mu$ by

$$\pi(\sigma_1, \ldots, \sigma_n) = \prod_{i=1}^n \mu_i(\sigma_i),$$

and for each $\sigma, \sigma' \in \Omega_\mu$, let

$$Pr_{\mu,i}(\sigma)(\sigma') = \begin{cases} \pi(\sigma')/\mu_i(\sigma_i) & \text{if } \sigma_i = \sigma_i' \\ 0 & \text{otherwise.} \end{cases}$$

Let $M_\mu = (\Omega_\mu, id_{\Omega_\mu}, \vec{Pr}_\mu)$. It is easy to check that $M_\mu$ is a $\Gamma$-structure; call it the **characteristic $\Gamma$-structure for** $\mu$. At each state in $M_\mu$, each player *i* is sure of his own strategy and has uncertainty about the strategies of his opponents; however, this uncertainty takes the form of a probability distribution weighted according to $\mu_{-i}$, so in effect each player *i* correctly ascribes the mixed strategy $\mu_j$ to each of his opponents $j \neq i$. It is well known (and easy to show) that a mixed strategy profile $\mu$ is a Nash equilibrium in the classical sense if and only if each player is rational (i.e. maximizing expected utility) at every state in the characteristic $\Gamma$-structure for $\mu$. Accordingly, we *define* a **Nash equilibrium** (in an $\mathcal{L}_B(\Phi_\Gamma)$-game) to be a mixed strategy profile $\mu$ such that $M_\mu \models RAT$. It is immediate that this definition generalizes the classical definition of Nash equilibrium.

We note that there are several other epistemic characterizations of Nash equilibrium besides the one presented here. While in the classical setting they all generate equivalent solution concepts, this need not be true in our more general model. We believe that investigating the solution concepts that arise by teasing apart these classically equivalent notions is an interesting and promising direction for future research.

In contrast to the classical setting, Nash equilibria are not guaranteed to exist in general; indeed, this is the case for the indignant altruism game of Example 2.

PROPOSITION 2. *There is no Nash equilibrium in the indignant altruism game.*

PROOF. We must show that for every mixed strategy profile
$$\mu = (\mu_A, \mu_B) \in \Delta(\{c, d\}) \times \Delta(\{c, d\}),$$
the corresponding characteristic $\Gamma$-structure $M_\mu \not\models RAT$.

Suppose first that $\mu_A(c) > 0$. Then $M_\mu \models \neg B_B \, play_A(d)$, which implies that Alice's utility at every state in $M_\mu$ coincides with the classical prisoner's dilemma, so she is not rational at any state where she cooperates. Since, by definition, $M_\mu$ contains a state where Alice cooperates, we conclude that $M_\mu \not\models RAT_A$, so $\mu$ cannot be a Nash equilibrium.

Suppose instead that $\mu_A(c) = 0$. Then $M_\mu \models B_B \, play_A(d)$, and so Alice, being sure of this, is not rational at any state where she defects, since by definition she is guaranteed a utility of $-1$ in that case. By definition, $M_\mu$ contains a state where Alice defects (in fact, Alice defects in every state), so we can conclude as above that $M_\mu \not\models RAT_A$, which means that $\mu$ cannot be a Nash equilibrium. □

What went wrong here? Roughly speaking, the utility functions in this game exhibit a kind of "discontinuity": the utility of defecting is $-1$ precisely when your opponent is 100% certain that you will defect. However, as soon as this probability dips below 100%, *no matter how small the drop*, the utility of defecting jumps up to at least 1.

Broadly, this issue arises in $\mathcal{L}$-games whenever $\mathcal{L}$ is limited to a coarse-grained notion of belief, such as the underlying language in this example, which only contains belief modalities representing 100% certainty. However, since coarseness is a central feature we wish to model, the lack of existence of Nash equilibria in general might be viewed as a problem with the notion of *Nash equilibrium* itself, rather than a defect of the underlying language. Indeed, the requirements that a mixed strategy profile must satisfy in order to qualify as a Nash equilibrium are quite stringent: essentially, each player must evaluate his choice of strategy *subject to the condition that his choice is common knowledge*! As we have seen, this condition is not compatible with rationality when a player's preference is to do something unexpected.

More generally, this tension arises with any solution concept that requires players to have common knowledge of the mixed strategies being played (the "conjectures", in the terminology of [2]). In fact, Proposition 2 relies only on second-order knowledge of the strategies: whenever Alice knows that Bob knows her play, she is unhappy. In particular, any alternative epistemic characterization of Nash equilibrium that requires such knowledge is subject to the same non-existence result. Furthermore, we can use the same ideas to show that there is no *correlated equilibrium* [1] in the indignant altruism game either (once we extend correlated equilibrium to our setting).

### 4.3 Rationalizability

In this section, we define rationalizability in language-based games in the same spirit as we defined Nash equilibrium in Section 4.2. As shown by Tan and Werlang [12] and Brandenburger and Dekel [4], common belief of rationality characterizes rationalizable strategies. Thus, we define rationalizability that way here.

Let $\mathcal{L}_{CB}(\Phi_\Gamma^{rat})$ be the language obtained by starting with the primitive propositions in $\Phi_\Gamma^{rat}$ and closing off under conjunction, negation, the modal operators $B_i$, for $i \in N$, and the modal operator $CB$. We read $CB\varphi$ as "there is common belief of $\varphi$". Extend $\llbracket \cdot \rrbracket_M$ to $\mathcal{L}_{CB}(\Phi_\Gamma^{rat})$ by setting

$$\llbracket CB\varphi \rrbracket_M \ := \ \bigcap_{k=1}^{\infty} \llbracket EB^k \varphi \rrbracket_M,$$

where

$$EB\varphi \ \equiv \ B_1\varphi \wedge \cdots \wedge B_n\varphi, \text{ and}$$
$$EB^k \varphi \ \equiv \ EB(EB^{k-1}\varphi).$$

For convenience, we stipulate that $EB^0 \varphi \equiv \varphi$. We read $EB\varphi$ as "everyone believes $\varphi$". Thus, intuitively, $CB\varphi$ holds precisely when everyone believes $\varphi$, everyone believes that everyone believes $\varphi$, and so on. We define a strategy $\sigma_i \in \Sigma_i$ to be **rationalizable** (in an $\mathcal{L}_B(\Phi_\Gamma)$-game) if the formula $play_i(\sigma_i) \wedge CB(RAT)$ is satisfiable in some $\Gamma$-structure.

Although there are no Nash equilibria in the indignant altruism game, as we now show, every strategy is rationalizable.

PROPOSITION 3. *Every strategy in the indignant altruism game is rationalizable.*

PROOF. Consider the $\Gamma$-structure in Figure 1.

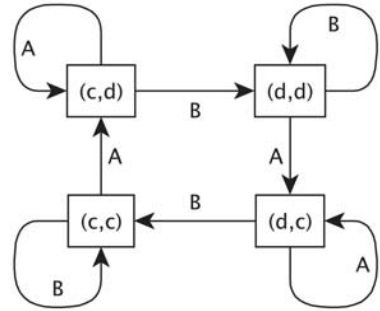

**Figure 1: A $\Gamma$-structure for indignant altruism.**

The valuations of the primitive propositions at each of the four states are labeled in the obvious way. Arrows labeled $i$ based at state $\omega$ point to all and only those states in $Pr_i[\omega]$ (so every probability measure has exactly one state in its support).

As discussed in Example 2, it is rational to cooperate in this game if you believe that your opponent believes that you will defect, and it is rational to defect if you believe that your opponent believes you will cooperate. Given this, it is not difficult to check that $RAT$ holds at each state of this $\Gamma$-structure, and therefore so does $CB(RAT)$. Thus, by definition, every strategy is rationalizable. □

Does every language-based game admit a rationalizable strategy? Every classical game does. This follows from the fact that every strategy in a Nash equilibrium is rationalizable, together with Nash's theorem that every (finite) game has a Nash equilibrium (cf. [10]). In the language-based setting, while it is immediate that every strategy in a Nash equilibrium is rationalizable, since Nash equilibria do not always exist, we cannot appeal to this argument. In fact, we have already seen an example of an $\mathcal{L}_B(\Phi_\Gamma)$-game that admits no rationalizable strategy.

PROPOSITION 4. *The deeply surprising proposal game has no rationalizable strategies.*

PROOF. Fix a $\Gamma$-structure $M = (\Omega, \vec{s}, \vec{Pr})$ and suppose for contradiction that $\omega \in \Omega$ is such that $\omega \models CB(RAT)$. Consider first the case where Alice does not *expect*$^*$ a proposal at state $\omega$, where "expect$^*$" denotes the infinitary notion of expectation at play in this example: for all $k \geq 0$, $\omega \models \neg P_A(P_B P_A)^k play_B(p)$. Thus, for all $k \geq 0$, $\omega \models B_A(B_B B_A)^k \neg play_B(p)$; taking $k = 0$, it follows that for all $\omega' \in Pr_A[\omega]$, $\omega' \models \neg play_B(p)$. Moreover, since $CB(RAT)$ holds at $\omega$, certainly $\omega' \models RAT_B$. But if Bob is rationally *not* proposing at $\omega'$, then he must at least consider it possible that Alice expects$^*$ a proposal: for some $k \in \mathbb{N}$, $\omega' \models P_B P_A (P_B P_A)^k play_B(p)$. But this implies that $\omega \models P_A(P_B P_A)^{k+1} play_B(p)$, contradicting our assumption. Thus, any state where $CB(RAT)$ holds is a state where Alice expects$^*$ a proposal.

So suppose that Alice expects$^*$ a proposal at $\omega$. It follows that there is some state $\omega'$ satisfying $\omega' \models play_B(p) \wedge CB(RAT)$. But if Bob is rationally playing $p$ at $\omega'$, there must be some state $\omega'' \in Pr_B[\omega']$ where Alice doesn't expect$^*$ it; however, we also know that $\omega'' \models CB(RAT)$, which we have seen is impossible.

This completes the argument: $CB(RAT)$ is not satisfiable. It is worth noting that this argument fails if we replace "expects$^*$" with "expects$^{\leq K}$", where this latter term is interpreted to mean

$$(\forall k \leq K)[\neg P_A(P_B P_A)^k play_B(p)].$$

□

In the full paper, we provide a condition that guarantees the existence of rationalizable strategies in $\mathcal{L}_B(\Phi_\Gamma)$-games. The essential ingredient is a kind of compactness assumption on the language $\mathcal{L}_B(\Phi_\Gamma^{rat})$. Roughly speaking, we require that no player can fail to be rational for an "infinitary" reason. All finitely-specified $\mathcal{L}_B(\Phi_\Gamma)$-games turn out to satisfy this condition, so we obtain the following:

THEOREM 1. *Every finitely-specified $\mathcal{L}_B(\Phi_\Gamma)$-game has a rationalizable strategy.*

Since we expect to encounter finitely-specified games most often in practice, this suggests that the games we are likely to encounter will indeed have rationalizable strategies.

## 5. ACKNOWLEDGEMENTS

Bjorndahl is supported in part by NSF grants IIS-0812045, CCF-1214844, DMS-0852811, and DMS-1161175, and ARO grant W911NF-09-1-0281. Halpern is supported in part by NSF grants IIS-0812045, IIS-0911036, and CCF-1214844, by AFOSR grant FA9550-08-1-0266, and by ARO grant W911NF-09-1-0281. Pass is supported in part by an Alfred P. Sloan Fellowship, a Microsoft Research Faculty Fellowship, NSF Awards CNS-1217821 and AFCF-1214844, NSF CAREER Award CCF-0746990, AFOSR YIP Award FA9550-10-1-0093, and DARPA and AFRL under contract FA8750-11-2-0211. The views and conclusions contained in this document are those of the authors and should not be interpreted as representing the official policies, either expressed or implied, of the Defense Advanced Research Projects Agency or the US Government.